\input amstex
\input amsppt.sty
\magnification=1200
\pageheight{24.5truecm}
\pagewidth{16truecm}
\NoRunningHeads
\loadbold
\frenchspacing

\def\pr{\partial}
\def\M{\operatorname{M}}
\def\Tr{\operatorname{Tr}}
\def\sfrac#1#2{{\textstyle\frac{#1}{#2}}}

\vglue 4cm
\topmatter

\title Legendre transformation for regularizable Lagrangians in field theory
\endtitle

\author
 Olga Krupkov\'a and Dana Smetanov\'a \\
{\eightpoint\it Mathematical Institute of the Silesian University in Opava\\
Bezru\v covo n\'am. 13, 746 01 Opava, Czech Republic
\\ e-mail:  Olga.Krupkova\@math.slu.cz, Dana.Smetanova\@math.slu.cz}
\endauthor

\abstract 
Hamilton equations based not only upon the Poincar\'e--Cartan equivalent of
a first-order Lagrangian, but rather upon its Lepagean equivalent are
investigated. Lagrangians which are singular within the
Hamilton--De Donder theory, but regularizable in this generalized sense are studied.
Legendre transformation for regularizable Lagrangians is proposed, and
Hamilton equations, equivalent with the Euler--Lagrange equations, are found. 
It is shown that all Lagrangians affine or quadratic in the 
first derivatives of the field variables are regularizable.
The Dirac field and the electromagnetic field are discussed in detail.

\noindent{\smc Keywords.}
Lagrangian, Poincar\'e--Cartan form, Lepagean form, 
Hamilton--De Donder equations, Hamilton $p_2$-equations, 
regularity, regularizable Lagrangian, Legendre transformation,
Dirac field, electromagnetic field

\noindent{\smc MS classification.}  
70G50, 58Z05

\noindent{\smc PACS numbers.} 02.40, 11.10

\endabstract

\thanks
Research supported by 
Grants MSM:J10/98:192400002, VS 96003 and FRV\v S
1467/2000 of the Czech Ministry of Education, Youth and Sports,
and Grant 201/00/0724 of the Czech Grant Agency
\endthanks

\endtopmatter

\bigskip
\heading 1. Introduction 
\endheading

If $\lambda$ is a Lagrangian defined on $J^1Y$ (the
first jet prolongation of a fibred manifold $\pi:Y \to X$), and $\theta_\lambda$ 
is its Poincar\'e--Cartan form then the {\it Euler--Lagrange equations} are 
equations for local sections $\gamma:X \to Y$ of $\pi$ as follows:
$J^1\gamma^* i_\xi d\theta_\lambda = 0$, for every vertical vector field $\xi$ 
on $J^1Y$.
A geometric setting for the Hamilton theory on fibred manifolds 
goes back to Goldschmidt and Sternberg, who in their famous paper [7] proposed
to consider {\it Hamilton equations}  as an extension of the Euler--Lagrange 
equations to local sections $\delta:X \to J^1Y$, namely, 
$\delta^* i_\xi d\theta_\lambda = 0$. To become {\it equivalent} with the 
Euler--Lagrange equations, the Lagrangian has to satisfy  the {\it regularity 
condition}
$$
\det \Bigl( \frac{\pr^2 L}{\pr y^\sigma_i \pr y^\nu_k}\Bigl) \ne 0.
\tag{1.1}
$$
Goldschmidt--Sternberg's approach, now adopted as 
standard (cf. [2], [5], [6], [8], [15-17] and many others), however seems not 
to be quite satisfactory. This turns out namely if generalizations to higher 
order are considered, or if concrete physical fields are studied: unfortunately, 
allmost all of them are degenerate in the sense of the regularity condition (1.1).  
Paralelly, Dedecker in 1977, and Krupka in 1983 considered another, from the 
mathematical point of view a more natural, extension of the Euler--Lagrange equations, 
based not upon a Poincar\'e--Cartan form as above, but rather upon a general 
{\it Lepagean equivalent of a Lagrangian} [3], [11].  As noticed already by 
Dedecker in [3], this approach opens a possibility to study ``regularizations" 
of singular Lagrangians.

This paper develops the idea to understand Hamilton equations in the above 
mentioned generalized sense. However, we differ from Dedecker and Krupka in some
points. First of all, we consider Lepagean equivalents which are 
{\it at most $2$-contact}, i.e. of the form $\theta_\lambda +$ some auxiliary 
$2$-contact term. Such Hamilton equations, first considered in [14] and called 
there {\it Hamilton $p_2$-equations} can be viewed as a ``first correction" to 
the standard Hamilton equations. In [14], we studied relations with the 
Euler--Lagrange equations, and obtained appropriate 
{\it regularity conditions}, generalizing (1.1). 
The aim of this paper is to propose {\it Legendre transformation for  
Hamilton $p_2$-equations}, and to apply the results to concrete physically 
interesting first order Lagrangians, namely, Lagrangians {\it affine} or 
{\it quadratic} in the first derivatives of the field variables. 
Comparing our approach with 
Dedecker [3], one can see that our concept of regularity is stronger, and 
Legendre transformation is understood in a completely different way. 

Contrary to the standard approach, where all affine and many
quadratic Lagrangians are singular, we show that all these Lagrangians are 
regularizable, admit Legendre transformation, and provide Hamilton equations 
which are {\it equivalent} with the Euler--Lagrange equations (i.e., do {\it not}
contain constraints). We also show that under certain additional conditions 
Hamilton $p_2$-equations of a Lagrangian coincide 
with the {\it usual} Hamilton equations of an approprite {\it equivalent Lagrangian}. 
We study in detail the case of the Dirac field and the electromagnetic field, 
and find the corresponding ``corrected" momenta and Hamiltonian which could be 
alternatively used for (unconstrained) quantization. 

Finally, we note that results and techniques presented in this paper 
can be generalized to higher-order variational problems [13], [18].

\heading 2. Preliminaries
\endheading

Throughout the paper all manifolds and mappings are smooth, and summation 
convention is used.
We consider a fibred
manifold $\pi:Y \to X$, $\dim X = n$, $\dim Y = m+n$, and its first (respectively, 
second) jet prolongation $\pi_1: J^1Y \to X$ (resp. $\pi_2: J^2Y \to X$). Natural 
fibred projections $J^kY \to J^lY$, where $0 \leq l < k \leq 2$,
are denoted by $\pi_{k,l}$.
A fibred chart on $Y$ (respectively, associated
chart on $J^1Y$) is denoted by $(V, \psi)$, $\psi =(x^i,y^\sigma)$ (respectively, 
$(V_1, \psi_1)$, where $V_1 = \pi_{1,0}^{-1}(V)$ and 
$\psi_1 =(x^i,y^\sigma,y^\sigma_i)$). We use the following notations:
$$
\omega_0 = dx^1\wedge dx^2\wedge \cdots \wedge dx^n,\quad
\omega_i = i_{\pr /\pr x^i} \omega_0,\quad
\omega_{ij} = i_{\pr / \pr x^j} \omega_i, \quad \cdots,
\tag{2.1}
$$
and
$$
\omega^\sigma = dy^\sigma - y^\sigma_j dx^j.
\tag{2.2}
$$
It is important to note that $(dx^i, \omega^\sigma, dy^\sigma_j)$ is a {\it basis
of $1$-forms on $J^1Y$}.
A mapping $\gamma:X \to Y$ defined on an open subset $U \subset X$ is 
called a {\it section} of the fibred manifold $\pi$ if the composite mapping
$\pi \circ \gamma$ is the identity mapping of $U$. Quite analogously, a section of the
fibred manifold $\pi_1$ is defined. Notice that a section of $\pi_1$
need not be of the form of prolongation of a section of $\pi$. Accordingly, 
a section $\delta$ of the fibred manifold $\pi_1$ is called {\it holonomic} if
$\delta = J^1\gamma$ for a section $\gamma$ of $\pi$. 

Recall that every $q$-form $\eta$ on $J^1Y$ admits a unique (canonical) decomposition
into a sum of $q$-forms on $J^2Y$ as follows:
$$
\pi^*_{2,1}\eta = h(\eta) + \sum^q_{k=1}p_k(\eta),
\tag{2.3}
$$
where $\pi_{2,1}$ is the canonical projection $J^2Y \to J^1Y$, $h(\eta)$ is a horizontal
form, called the horizontal part of $\eta$, and $p_k (\eta)$, $1 \leq k \leq q$, is a
$k$-contact form, called the $k$-contact part of $\eta$ (see e.g. [9]).
For our purposes it is sufficient to recall that in fibred coordinates a horizontal
form on $J^1Y$ is expressed by means of wedge products of the differentials 
$dx^i$ only, with the components dependent upon the coordinates 
$(x^i,y^\sigma,y^\sigma_j)$. Similarly, a $1$-contact (respectively, $2$-contact)
form contains only wedge products of the differentials $dx^i$ with {\it one} 
(respectively, {\it two}) of the contact forms (2.2).

By a {\it first-order Lagrangian} we shall mean a horizontal {\it n}-form $\lambda$ on 
$J^1Y$. This means that in every fibred chart, 
$$
\lambda = L \omega_0
\tag{2.4}
$$
where $L = L(x^i,y^\sigma,y^\sigma_j)$.
A form $\rho$ is called a {\it Lepagean equivalent} of a Lagrangian $\lambda$ 
if (up to a projection) $h(\rho)=\lambda$, and $p_1(d\rho)$ is a 
$\pi_{2,0}$-horizontal 
form [9]. All (first order) Lepagean equivalents of a Lagrangian of order one take 
the form
$$
\rho = \theta_\lambda + \nu,
\tag{2.5}
$$
where $\theta_\lambda$ is the {\it Poincar\'e--Cartan equivalent} of $\lambda$, i.e.
$$
\theta_\lambda = L \omega_0 + \frac{\pr L}{\pr y^\sigma_j} \omega^\sigma \land
\omega_j,
\tag{2.6}
$$ 
and $\nu$ is an {\it arbitrary at least $2$-contact}
$n$-form, i.e. such that $h(\nu) = p_1(\nu) = 0$.

If $\rho$ is a Lepagean equivalent of $\lambda$ then the $(n+1)$-form
$E_\lambda = p_1 (d\rho)$ is called the {\it Euler--Lagrange form} of 
the Lagrangian $\lambda$. Two Lagrangians $\lambda_1$ and $\lambda_2$ are called
{\it equivalent} if $E_{\lambda_1} = E_{\lambda_2}$. Recall that 
Lagrangians $\lambda_1$ and $\lambda_2$ are equivalent on an open set 
$U \subset J^1Y$ if and only if there exists an 
$(n-1)$-form $\varphi$ such that $\lambda_2 = \lambda_1 + h(d\varphi)$ [9].

Besides the Poincar\'e--Cartan equivalent (2.6), the family (2.5) of Lepagean 
equivalents of a Lagrangian contains another distinguished Lepagean
equivalent uniquely determined by the Lagrangian, namely,
$$
\rho_\lambda = L\,\omega_0 + \sum_{k=1}^n \, \left(\sfrac{1}{k!} \right)^2
\frac{\pr^k L}
{\pr y^{\sigma_1}_{j_1} \cdots \pr y^{\sigma_k}_{j_k}} \,\omega^{\sigma_1} \land 
\dots \land \omega^{\sigma_k}
\land \omega_{j_1 \cdots j_k}
\tag{2.7}
$$
[10] (cf. also [1]). It is called the {\it Krupka equivalent} of $\lambda$, and
has the following important property: 
{\it $d\rho_\lambda = 0$ if and only if $E_\lambda = 0$}; the latter condition,
however, means that $\lambda = h(d\varphi)$ (a so called trivial Lagrangian).

With the help of Lepagean equivalents of a Lagrangian one obtains an intrinsic 
formulation of the {\it Euler--Lagrange} and {\it Hamilton equations} as follows  
(cf. [9], [11]).
A section $\gamma $ of the fibred manifold $\pi$ is an {\it extremal} of $\lambda$ 
if and only if
$$
J^1\gamma^*i_{J^1\xi}d\rho = 0
\tag{2.8}
$$
for every $\pi$-vertical vector field $\xi$ on $Y$.
A section $\delta$ of the fibred manifold $\pi_1$ is called  a
{\it Hamilton extremal} of $\rho$ if 
$$
\delta^*i_\xi d\rho = 0,
\tag{2.9}
$$
for every $\pi_1$-vertical vector field $\xi$ on $J^1Y$. The equations (2.8) and
(2.9) are called the {\it Euler--Lagrange} and the {\it Hamilton equations}, 
respectively.

Notice that while the Euler--Lagrange equations (2.8) are uniquely determined by the 
Lagrangian,
Hamilton equations (2.9) depend upon the choice of $\nu$. Consequently, one has many 
different ``Hamilton theories" associated to a given variational problem. 

Clearly, if $\gamma$ is an extremal then $J^1\gamma $ is a Hamilton extremal; 
conversely, however, a Hamilton extremal need not be holonomic, and thus a jet 
prolongation of some extremal.
This suggests a definition of regularity as follows:
A Lepagean form is called {\it regular} if every its Hamilton extremal is
holonomic [12].

Hamilton equations (2.9) where $\rho = \theta_\lambda$ (respectively, 
$\rho$ is {\it at most} $2$-contact) are called 
{\it Hamilton-De Donder equations} [4], [7] (respectively, 
{\it Hamilton $p_2$-equations} [14]).

\bigskip
\heading 3. Hamilton $p_2$-equations and Legendre transformation
for first-order Lagrangians
\endheading

In the sequel we consider Lepagean forms (2.5) where $\nu$ is {\it $2$-contact}.
Moreover, we suppose $\nu = p_2(\beta)$, where $\beta$ is defined on $Y$ and 
such that $p_i(\beta) = 0$ for all $i \geq 3$.
Hence, in fibred coordinates
$$
\rho = 
 L\omega_0 + {\pr L\over \pr y^\sigma_j}\omega^\sigma \wedge \omega_j +
g^{ij}_{\sigma \nu}\, \omega^\sigma \wedge \omega^\nu \wedge \omega_{ij},
\tag {3.1}
$$
where the functions $g^{ij}_{\sigma\nu}$ do not depend upon the $y^\kappa_l$'s and
satisfy  the conditions
$$
g^{ij}_{\sigma \nu} = -g^{ij}_{\nu \sigma},\quad
g^{ij}_{\sigma \nu} = -g^{ji}_{\sigma \nu},\quad
g^{ij}_{\sigma \nu} = g^{ji}_{\nu \sigma}.
\tag{3.2}
$$
Note that (3.2) mean that only
$$
{m \choose 2} \cdot {n \choose 2} = \sfrac{1}{4} mn\,(m-1)(n-1)
$$
of the $mn \times mn$ functions $g^{ij}_{\sigma \nu}$ are independent.

\proclaim{Theorem 1 \rm[14]}
Let $\lambda$ be a first-order Lagrangian, $\rho $ its Lepagean equivalent 
as above.
Assume that the matrix
$$
\left( {{\pr ^2L} \over {\pr y^\sigma_i \pr y^\nu_j}} -
4g^{ij}_{\sigma \nu}\right)
\tag{3.3}
$$ 
with rows (respectively, columns) labelled by the pair $(\sigma,i)$ 
(respectively, $(\nu,j)$), is regular. 
Then $\rho$ is regular. Moreover, every Hamilton extremal $\delta$ of $\rho$ is of
the form $\delta = J^1\gamma$, where $\gamma$ is an extremal of $\lambda$.
\endproclaim

The proof is obtained by a direct calculation from (2.9), and can be found in [14].

In view of the above theorem we have the following concept: 

\definition{Definition 1}
Let $W \subset J^1Y$ be an open set.
A Lagrangian $\lambda$ is called {\it regularizable on $W$} if it has a regular 
Lepagean equivalent $\rho$ (3.1) defined on $W$. If $W = J^1Y$ we say that 
$\lambda$ is {\it globally regularizble}. We say that $\lambda$ is {\it locally
regularizable} if it is regularizable in a neighbourhood of every point in $J^1Y$.
The corresponding Lepagean equivalent $\rho$ is then called
a {\it (local) regularization} of $\lambda$. 
\enddefinition

Note that for regularizable Lagrangians, the problem of solving the 
Euler--Lagrange equations is {\it equivalent} to the problem of solving 
(appropriate) Hamilton equations.

An important class of regularizable Lagrangians is characterized by the 
following proposition.

\proclaim{Proposition 1}
Let $m \geq2$. Then
every Lagrangian $L$ such that 
$$
L = a + b_\sigma^j y^\sigma_j + c_{\sigma \nu}^{jk} y^\sigma_j y^\nu_k,
\tag{3.4}
$$
where $a$, $b_\sigma^j$ and $c_{\sigma \nu}^{jk}$ are functions of $(x^i, y^\rho)$,
is locally regularizable.

In particular,
every first-order Lagrangian affine (respectively, quadratic) in the $y^\rho_p$'s 
is locally regularizable.
\endproclaim

\demo{Proof}
By assumption, 
$\pr^2 L/\pr y^\sigma_i \pr y^\nu_j$ are functions defined on an open 
subset of the total space $Y$.
Let $\det(\pr^2 L/\pr y^\sigma_i \pr y^\nu_j) = 0$
at a point $x \in Y$. Since $m > 2$, one can find functions 
$g^{ij}_{\sigma \nu}$, antisymmetric in $(\sigma \nu)$ and $(ij)$, defined in 
a neighbourhood  of $x$ and such that at $x$ the condition (3.3) is satisfied. However,
since the determinant is a continuous function, the corresponding matrix must be
nondegenerate in a neighbourhood $U$ of $x$. With these $g$'s (independent of 
the $y^\rho_p$'s, as desired), the form 
$\rho = \theta_\lambda + g^{ij}_{\sigma \nu} \omega^\sigma \land \omega^\nu \land 
\omega_{ij}$ is a regularization of $\lambda$ on $W = \pi_{1,0}^{-1}(U)$.
\qed
\enddemo

The following theorem provides a {\it generalization of 
Legendre transformation to} singular in the
standard sense, but {\it regularizable Lagrangians}.

\proclaim{Theorem 2} Consider a Lepagean form $\rho$ given in a fibred chart
$(V,\psi)$, $\psi = (x^i,y^\sigma)$  by (3.1) (3.2). 
Put
$$
p^i_\sigma = {{\pr L} \over {\pr y^\sigma_i}} - 4g^{ij}_{\sigma \nu}y^\nu_j.
\tag {3.5}
$$
Let $x \in V_1 \subset J^1Y$ be a point. If the matrix (3.3) is regular in a 
neighbourhood $W \subset V_1$ of $x$, 
then $(x^i, y^\sigma,  y^\sigma_j)  \rightarrow (x^i, y^\sigma,  p^j_\sigma)$
is a coordinate transformation on $W$.
\endproclaim

\demo{Proof}
The above theorem follows immediately from the fact that
$$
\frac{\pr {p^i_\sigma}}{\pr y^\nu_j} =
\frac {\pr ^2L}{\pr y^\sigma_i \pr y^\nu_j} -
 4g^{ij}_{\sigma \nu}.
$$
\qed
\enddemo

Denote
$$
H = -L + {{\pr L} \over {\pr y^\sigma_i}}y^\sigma_i -
2g^{ij}_{\sigma \nu} y^\sigma_i y^\nu_j
 = -L + p^i_\sigma y^\sigma_i +
2g^{ij}_{\sigma \nu} y^\sigma_i y^\nu_j.
\tag {3.6}
$$
Now the Lepagean form (3.1), (3.2) reads
$$
\aligned
\rho &= L\omega_0 + {\pr L\over \pr y^\sigma_j}\omega^\sigma \wedge \omega_j +
 g^{ij}_{\sigma \nu}\, \omega^\sigma \wedge \omega^\nu \wedge \omega_{ij} 
= \Bigl(L - {{\pr L} \over {\pr y^\sigma_i}}y^\sigma_i 
 + 2g^{ij}_{\sigma \nu} y^\sigma_i y^\nu_j \Bigr)\omega_0 \\
 &+ \Bigl({{\pr L} \over {\pr y^\sigma_i}} - 4g^{ij}_{\sigma \nu}y^\nu_j \Bigr)
 {dy}^\sigma \wedge \omega_i +
 g^{ij}_{\sigma \nu}\, {dy}^\sigma \wedge {dy}^\nu \wedge \omega_{ij} \\
&= - H \omega_0 + p^i_\sigma{dy}^\sigma \wedge \omega_i +
 g^{ij}_{\sigma \nu}\, {dy}^\sigma \wedge {dy}^\nu \wedge \omega_{ij}.
\endaligned
\tag{3.7}
$$

In analogy with the standard terminology we shall call $H$ 
the {\it Hamiltonian} and $p^i_\sigma$ {\it momenta} of
the Lepagean form $\rho$ (3.1), (3.2), and the corresponding coordinate 
transformation {\it Legendre transformation}; accordingly, the coordinates
$(x^i, y^\sigma,  p^i_\sigma)$ will be referred to as
{\it Legendre coordinates} of $\rho$.

\proclaim{Corollary 1}
Let $(x^i, y^\sigma,  p^i_\sigma)$ be the Legendre transformation associated 
with a Lepagean form $\rho$ (3.1), (3.2).
Then the matrix
$$
\left ( {\pr^2 H \over \pr p^i_\sigma \pr p^j_\nu} \right)
\tag{3.8}
$$
is regular and inverse to the matrix (3.3).
\endproclaim

\demo {Proof}
Explicit computations lead to
$$
\aligned
{\pr L \over \pr y^\nu_j} &= p^j_\nu + 4 g^{jk}_{\nu \kappa}\, y^\kappa_k,
\quad
{\pr L \over \pr p^i_\sigma} =  {\pr L \over \pr y^\nu_j}
{\pr y^\nu_j \over \pr p^i_\sigma} = 
\left(p^j_\nu + 4 g^{jk}_{\nu \kappa}\, y^\kappa_k\right) 
{\pr y^\nu_j \over \pr p^i_\sigma},
\\
{\pr H \over \pr p^i_\sigma}  &= -{\pr L \over \pr p^i_\sigma}
   + y^\sigma_i + p^j_\nu {\pr y^\nu_j \over \pr p^i_\sigma}
   +  4 g^{kj}_{\kappa \nu} y^\kappa_k {\pr y^\nu_j \over \pr p^i_\sigma}
= y^\sigma_i,\quad
{\pr^2 H \over \pr p^i_\sigma \pr p^j_\nu}  =  {\pr y^\sigma_i \over \pr p^j_\nu},
\endaligned
$$
proving the assertion.
\qed  
\enddemo

Expressing Hamilton $p_2$-equations (2.9) in Legendre coordinates we get
$$
\aligned
&{{\pr H} \over {\pr y^\sigma}} = - {{\pr p^i_\sigma} \over {\pr x^i}} \ + \ 
 4{{\pr g^{ij}_{\sigma \nu}} \over {\pr x^j}}{{\pr y^\nu} \over {\pr x^i}}
 \ + \ 2 \left ({{\pr g^{ij}_{\kappa \nu}} \over {\pr y^\sigma}}\ +\ 
 {{\pr g^{ij}_{\sigma \kappa}} \over {\pr y^\nu}} \ + \ 
 {{\pr g^{ij}_{\nu \sigma}} \over {\pr y^\kappa}}\right )
 {{\pr y^\kappa} \over {\pr x^i}} {{\pr y^\nu} \over {\pr x^j}}, \\
&{{\pr H} \over {\pr p^i_\sigma}} = {{\pr y^\sigma} \over {\pr x^i}},
\endaligned
\tag {3.9}
$$
or, equivalently,
$$
\aligned
&{{\pr H} \over {\pr y^\sigma}} = - {{\pr p^i_\sigma} \over {\pr x^i}} \ + \ 
 4{{\pr g^{ij}_{\sigma \nu}} \over \pr x^j}{{\pr H} \over {\pr p^i_\nu}} \ + \ 
 2 \left ({{\pr g^{ij}_{\kappa \nu}} \over {\pr y^\sigma}}\ +\ 
 {{\pr g^{ij}_{\sigma \kappa}} \over {\pr y^\nu}} \ + \ 
 {{\pr g^{ij}_{\nu \sigma}} \over {\pr y^\kappa}}\right )
 {{\pr H} \over {\pr p^i_\kappa}}{{\pr H} \over {\pr p^j_\nu}}, \\
&{{\pr H} \over {\pr p^i_\sigma}} = {{\pr y^\sigma} \over {\pr x^i}}.
\endaligned
\tag{3.10}
$$

\proclaim{Corollary 2}
If the $n$-form
$$
\eta = g^{ij}_{\sigma \nu}\, {dy}^\sigma \wedge {dy}^\nu \wedge \omega_{ij}
\tag {3.11}
$$
is closed then the above Hamilton $p_2$-equations take the form
$$
{{\pr H} \over {\pr y^\sigma}} = - {{\pr p^i_\sigma} \over {\pr x^i}}, \quad
{{\pr H} \over {\pr p^i_\sigma}} = {{\pr y^\sigma} \over {\pr x^i}}.
\tag{3.12}
$$
\endproclaim

\definition{Remark 1}
Compared with Dedecker [3], we differ in both the 
definition of regularity and Legendre transformation. Dedecker's regularity is
weaker---Hamilton equations regular in his sense need not be equivalent with the 
Euler--Lagrange equations. Also, Legendre transformation is  
completely different: while Dedecker's Legendre transformation is a map to
a certain new space of higher dimension than that of the dynamical space, with 
unclear relations to regularity and to Lagrangian dynamics,
Legendre transformation proposed above has similar properties as the Legendre 
transformation in classical mechanics (if the Lagrangian is regular in our sense, Legendre
transformation becomes a local diffeomorphism of the space where the 
dynamics proceeds, providing a ``canonical" form of the motion equations).
Some further geometric properties of this Legendre transformation are clarified 
in [13].
\enddefinition

\heading 4. Satellite Lagrangians
\endheading

We shall investigate the meaning of the $2$-contact term in the Lepagean
equivalent $\rho$ (3.1) of a Lagrangian. Keeping notations introduced so far,
we start with the following interesting assertions:

\proclaim{Lemma 1}
It holds (up to the projection $\pi_{1,0}$)
$$
\eta = \rho_{h(\eta)}.
\tag{4.1}
$$
\endproclaim

\demo{Proof}
Denote $h(\eta) = l\, \omega_0$; then obviously,
$$
l = 2 g^{ij}_{\sigma \nu}\, y^\sigma_i y^\nu_j,
\tag{4.2}
$$
and we obtain using (2.7) and (2.2)
$$
\aligned
\rho_{h(\eta)} &= l\,\omega_0 + \frac{\pr l}{\pr y^\sigma_i} \,\omega^\sigma 
\land  \omega_i +
\sfrac{1}{4} \frac{\pr^2 l}
{\pr y^\sigma_i\pr y^\nu_j} \,\omega^\sigma \land \omega^\nu \land \omega_{ij} \\
&= 2 g^{ij}_{\sigma \nu}\, y^\sigma_i y^\nu_j \,\omega_0 
+ 4 g^{ij}_{\sigma \nu}\,y^\nu_j \, \omega^\sigma \land  \omega_i
+ g^{ij}_{\sigma \nu}\, \omega^\sigma \land \omega^\nu \land \omega_{ij} \\
&= g^{ij}_{\sigma \nu}\,dy^\sigma \land dy^\nu\land \omega_{ij} = \eta.
\endaligned
\tag{4.3}
$$
\qed
\enddemo

\proclaim{Lemma 2}
Put
$$
\bar \lambda = \lambda - h(\eta) = h(\rho - \eta) , \quad \text{i.e.} \quad
\bar L = L - l.
\tag{4.4}
$$
Then
$$
\rho = \theta_{\bar \lambda} + \rho_{h(\eta)}, \quad \text{i.e.} \quad
\theta_{\bar \lambda} = \rho - \eta.
\tag{4.5}
$$
\endproclaim

\demo{Proof}
Indeed, by (4.3) and (3.1),
$\rho_{h(\eta)} = \theta_{h(\eta)} + p_2(\rho)$, hence, by (4.4),
$$
\rho = \theta_\lambda + p_2(\rho) = \theta_\lambda  - 
\theta_{h(\eta)} + \rho_{h(\eta)} = 
\theta_{\bar \lambda} + \rho_{h(\eta)}.
$$
\qed
\enddemo

Note that by (4.5) and (3.7), 
$$
\theta_{\bar \lambda} = - H\omega_0 + p_\sigma^i dy^\sigma \land \omega_i.
\tag{4.6}
$$

Let us denote by $\widetilde p^j_\sigma(L)$ and $\widetilde H(L)$ the
De Donder momenta and Hamiltonian associated with 
a Lagrangian $L$. Recall that 
$$
{\widetilde p^i_\sigma}(L) = {\pr L \over \pr y^\sigma_i},\quad
{\widetilde H}(L) = - L + {\widetilde p^i_\sigma}(L)\, y^\sigma_i
\tag{4.7}
$$
[4], [7]. Note that for our momenta and Hamiltonian (3.5) and (3.6) we
obtain
$$
p_\sigma^i = {\widetilde p^i_\sigma}(L) - {\widetilde p^j_\sigma}(l),
\quad
H = {\widetilde H}(L) - {\widetilde H}(l) = {\widetilde H}(L) - l.
\tag{4.8}
$$
Consequently, we have the following lemma.

\proclaim{Lemma 3}
It holds
$$
H = {\widetilde H}(\bar L), \quad  
p_\sigma^i = {\widetilde p^i_\sigma}(\bar L).
\tag{4.9}
$$
Moreover, the regularity condition (3.3) is equivalent with the ``standard" 
regularity condition (1.1) for $\bar L$, i.e., with
$$
\det \Bigl( \frac{\pr^2 \bar L}{\pr y^\sigma_i \pr y^\nu_k}\Bigl) \ne 0.
\tag{4.10}
$$
\endproclaim

In view of the above results we shall call the Lagrangian 
$h(\eta)$ a {\it satellite} of $\lambda$, and
the Lagrangian $\bar \lambda$ a {\it dedonderization} of $\lambda$.

Now, from Lemma 1 and 2 we immediately obtain the following important result:

\proclaim{Proposition 2}
If the form $\eta$ is closed then the Lagrangian $\bar \lambda$ is equivalent 
with $\lambda$, and $d\rho = d\theta_{\bar \lambda}$.
\endproclaim

In Section 3 we introduced {\it regularization} as a procedure to find for a 
Lagrangian appropriate {\it Hamilton $p_2$-equations} (i.e., a certain
``correction" to the Hamilton De-Donder equations of $L$) which are
{\it equivalent} with the Euler--Lagrange equations, hence represent a
suitable (unconstrained) alternative for solving the extremal problem.
Now, taking into account all the above properties of satellite Lagrangians
we conclude that regularization can be understood also {\it in a different way} 
as a procedure {\it to find to a given Lagrangian an appropriate satellite
in such a way that the Hamilton--De Donder equations of the new Lagrangian
would be equivalent with the Euler--Lagrange equations}:

\proclaim{Theorem 3}
Let $\lambda$ be a regularizable Lagrangian. Then for every its (local) 
regularization $\rho$ such that $d\eta = 0$, the Lagrangian
$\bar \lambda = \lambda - h(\eta)$ is equivalent with $\lambda$, satisfies 
the ``standard" regularity condition (4.10), and the Hamilton $p_2$-equations 
of $\lambda$ based upon $\rho$ coincide 
with the Hamilton--De Donder equations of the Lagrangian $\bar \lambda$. 
\endproclaim

\heading 5. Examples of Legendre transformations for first-order Lagrangians
\endheading

The above results can be directly applied to concrete Lagrangians. Let us consider
two important cases: {\it Lagrangians affine in the first derivatives of the 
field variables} (in particular, 
the {\it Dirac field}), and the {\it electromagnetic field}.

\medskip
\noindent
{\bf 5.1. Affine Lagrangians.}
Recall that by Proposition 1, if the fibre dimension $m$ is at least $2$,
all Lagrangians affine in the first derivatives 
are locally regularizable and admit Legendre transformation introduced in 
Section 3. Assume
$$
L = L_0 + L^i_\sigma\, y^\sigma_i
\tag{5.1}
$$
where $L_0$ and $L^i_\sigma$ ($1 \leq i \leq n$, $1 \leq \sigma \leq m$)
are functions of $(x^j,\, y^\nu)$.
Note that the De Donder momenta and Hamiltonian of (5.1) take the
form $\widetilde p^j_\sigma = L^j_\sigma$, $\widetilde H = - L_0$, i.e.,
they are defined on an open subset of the {\it total space} $Y$, and the 
corresponding Hamilton equations
must be treated as {\it constrained}, within the range of the Dirac theory 
of constrained systems (cf. eg. [6]).
On the other hand, we get by (3.5) and (3.6),
$$
p_\sigma^i = L_\sigma^i - 4g^{ij}_{\sigma\nu} y^\nu_j, \quad
H = - L_0 - 2g^{ij}_{\sigma\nu} y^\sigma_i y^\nu_j,
\tag{5.2}
$$
where $(g^{ij}_{\sigma\nu})$ is a regular $(mn \times mn)$-matrix.
We can see that the domain
of definition of the functions (5.2) is an open subset of $J^1Y$;
momenta are functions {\it affine} in the  $y^\nu_j$'s, and $H$ 
(in Legendre coordinates) is a {\it polynomial of degree $2$} in momenta.  
By Theorem 3, we should choose the $g^{ij}_{\sigma\nu}$ in such a 
way that the form $\eta$ (3.11) be closed. Then the corresponding satellite
Lagrangian is trivial, and Hamilton equations take the form (3.12). 

Let us discuss in more detail the case $m=2$, and $n=2$ (respectively, $n=4$).

\smallskip
(i) $n = \dim X = 2$.
The conditions (3.2) on the $g^{ij}_{\sigma\nu}$'s mean
that only one of these functions is independent, say, $g^{12}_{12}$.
Denote $u = 4g^{12}_{12}$, and assume $u \ne 0$. The condition $d\eta = 0$, 
i.e., $du \land dy^1 \land dy^2 = 0$ means that $u = u(y^1, y^2)$.

As above, we consider Lepagean equivalents of the Lagrangian (5.1) 
in the form
$$
\rho = \left (L_0\, +\, L^i_\sigma y^\sigma_i \right)\, dx^1 \land dx^2 + 
L^i_\sigma\, \omega^\sigma \land \omega_i
+ u\  \omega^1 \land \omega^2
\tag{5.3}
$$
(where summation runns through $1 \leq i,\,\sigma \leq 2$).
The regularity condition (3.3) reads
$$
\det \pmatrix 
0&0&0&-u \\
0&0&u&0\\
0&u&0&0\\
-u&0&0&0\\
\endpmatrix \ne 0,
\tag{5.4}
$$
and is clearly satisfied.
Momenta become
$$
p^1_1 =  L^1_1 - u y^2_2, \quad
p^1_2 = L^1_2 + u y^1_2 ,\quad
p^2_1 = L^2_1 + u y^2_1, \quad
p^2_2 = L^2_2 - u y^1_1.
\tag{5.5}
$$
Since the inverse to the Legendre transformation takes the form
$$
y^1_1 = {1 \over u} \left ( L^2_2 - p^2_2 \right), \,\,
y^2_2 = {1 \over u} \left ( L^1_1 - p^1_1 \right), \,\, 
y^2_1 = - {1 \over u} \left (L^2_1 - p^2_1 \right), \,\, 
y^1_2 = - {1 \over u} \left (L^1_2 - p^1_2 \right), 
\tag{5.6}
$$
the Hamiltonian in the Legendre coordinates reads
$$
H = - L_0 + {1 \over u} \Bigl( L^1_2 L^2_1 - L^1_1 L^2_2 + 
 p^1_1 L^2_2  + p^2_2 L^1_1  - p^1_2 L^2_1 - p^2_1 L^1_2
 - p^1_1 p^2_2
 + p^1_2 p^2_1 \Bigr).
\tag {5.7}
$$
Hamilton $p_2$-equations 
in the Legendre coordinates take the form 
$$
\aligned
&{{\pr H} \over {\pr y^1}} = - {{\pr p^1_1} \over {\pr x^1}}
- {{\pr p^2_1} \over {\pr x^2}}, \quad
{{\pr H} \over {\pr y^2}} = - {{\pr p^1_2} \over {\pr x^1}}
- {{\pr p^2_2} \over {\pr x^2}}; \\
&{{\pr H} \over {\pr p^1_1}} = {{\pr y^1} \over {\pr x^1}}, \quad
{{\pr H} \over {\pr p^1_2}} = {{\pr y^2} \over {\pr x^1}}, \quad
{{\pr H} \over {\pr p^2_1}} = {{\pr y^1} \over {\pr x^2}}, \quad
{{\pr H} \over {\pr p^2_2}} = {{\pr y^2} \over {\pr x^2}}.
\endaligned
\tag{5.8}
$$
Note that for every fixed $u \ne 0$ we have obtained Hamilton equations 
{\it equivalent} with the Euler-Lagrange equations.

As an illustration, let us consider the {\it Dirac field}. In this case we have
$X = R^2$, $Y = R^2 \times R^2$, i.e. $J^1Y = R^2 \times R^2 \times R^4$, with the 
global coordinates denoted by $(x^\mu, \psi, \bar \psi, \pr_\mu \psi, 
\pr_\mu \bar \psi)$, $\mu = 1,2$, Lagrangian $L$ takes the form
$$
L = \sfrac{i}{2} (\bar \psi \gamma^\mu \pr_\mu \psi +
 \pr_\mu \bar \psi \gamma^\mu \psi) - \bar \psi m \psi,
\tag{5.9}
$$
and it is apparently degenerate in the sense of the standard regularity condition
(1.1). However, for every nonzero function $u(\psi, \bar \psi)$, the form $\rho$ 
(5.3) is a {\it global regularization} of $L$. Computing the corresponding 
{\it satellite Lagrangian} for (5.9) we get
$$
l = - u\, \epsilon^{\mu \nu} \pr_\mu \bar \psi \, \pr_\nu \psi,
\tag{5.10}
$$
where $\epsilon^{\mu \nu}$ is the Levi-Civita symbol. Now the Lagrangian
$$
\bar L = \sfrac{i}{2} (\bar \psi \gamma^\mu \pr_\mu \psi +
 \pr_\mu \bar \psi \gamma^\mu \psi) - \bar \psi m \psi +
u\, \epsilon^{\mu \nu} \pr_\mu \bar \psi \, \pr_\nu \psi
\tag{5.11}
$$
is a dedonderization of the Dirac field Lagrangian, which is {\it  regular}
in the standard sense. Accordingly, Hamilton--De Donder equations of (5.11)
are equivalent with the Dirac field equations, and in this sense, the (standard)
De Donder Hamiltonian and momenta of (5.11) (which, however, are precisely
the functions $H$ and $p$'s given by (5.7) and (5.5)), should represent possibly 
a better physical alternative for (unconstrained) quantization of the Dirac field.
Explicitly, the ``new" Hamiltonian reads by (4.8)
$$
H = u\, \epsilon^{\mu \nu}\pr_\mu \bar \psi\, \pr_\nu \psi + \bar \psi m \psi.
\tag{5.12}
$$

Note that in the above formulas the most simple admissible choice is $u \ne 0$ a 
{\it constant} function.

\smallskip
(ii) $n = \dim X = 4$. 
In this case we get $6$ independent functions 
$g^{ij}_{\sigma\nu}$. Denote
$$
u_1 = 4g^{12}_{12}, \,\, 
u_2 = 4g^{13}_{12}, \,\,
u_3 = 4g^{14}_{12}, \,\,
u_4 = 4g^{23}_{12}, \,\,
u_5 = 4g^{24}_{12}, \,\,
u_6 = 4g^{34}_{12}.
\tag{5.13}
$$
The matrix (3.3) takes the form
$$
\pmatrix 
0&-\M\\
\M&0
\endpmatrix,
\quad \text{where} \quad
M = \pmatrix
0&u_1&u_2&u_3\\
-u_1&0&u_4&u_5\\
-u_2&-u_4&0&u_6\\
-u_3&-u_5&-u_6&0
\endpmatrix.
\tag{5.14}
$$
We can see that for any choice of functions $u_k(x^i,y^\sigma)$,
$1 \leq k \leq 6$, such that $\det \M \ne 0$ we obtain a regular
Hamilton $p_2$-theory, based upon the Lepagean form
$$
\aligned
\rho &= \left ( L_0\, +\, L^i_\sigma y^\sigma_i \right )\,dx^1 
\land dx^2 \land dx^3 \land dx^4 +
L^i_\sigma\, \omega^\sigma \land \omega_i\\
&+ u_1\, \omega^1 \land \omega^2 \land \omega_{12} 
+ u_2\, \omega^1 \land \omega^2 \land \omega_{13}
+ u_3\, \omega^1 \land \omega^2 \land \omega_{14}\\
&+ u_4\, \omega^1 \land \omega^2 \land \omega_{23}
+ u_5\, \omega^1 \land \omega^2 \land \omega_{24}
+ u_6\, \omega^1 \land \omega^2 \land \omega_{34}.
\endaligned
\tag{5.15}
$$
Expressing the corresponding {\it satellite Lagrangian} for the Dirac field
explicitly, we easily obtain
$$
l = - \sum_{\mu,\nu} u_{(\mu, \nu)}\, \epsilon^{\mu \nu} \, \pr_\mu \bar \psi\,
\pr_\nu \psi
\tag{5.16}
$$
where $u_{(\mu, \nu)} = u_{(\nu, \mu)}$ and the notation $u_1 = u_{(1,2)}$,
$u_2 = u_{(1,3)}$, $u_3 = u_{(1,4)}$, $u_4 = u_{(2,3)}$, $u_5 = u_{(2,4)}$,
$u_6 = u_{(3,4)}$ is used. ``Corrected" momenta can now be obtained by a short 
routine calculation from (3.5), and the Hamiltonian takes by (4.8) the form 
$H = l + \bar \psi m \psi$. In comparison with the usual formulas they
differ by additional terms---(De Donder) momenta and Hamiltonian of the
satellite (5.16) of the Lagrangian (5.9).

Note that on the fibred manifold $R^4 \times R^2 \to R^4$ the most simple 
{\it global} regularization of the Dirac Lagrangian is obtained for
$u_i$, $1 \leq i \leq 6$, constant functions; by (5.13) some of them may even
equal zero (a simple choice is e.g. $u_3, u_4 \ne 0$, 
$u_1 = u_2 = u_5 = u_6 = 0$).

\medskip
\noindent
{\bf 5.2. The electromagnetic field}.
For the electromagnetic field Lagrangian
$$
L = - \frac{1}{4} F_{\mu\nu}F^{\mu\nu} =  \frac{1}{2} 
(y^\sigma_\nu y^\nu_\sigma - {\frak g}^{\sigma\nu} {\frak g}_{\mu\rho}\, 
y^\mu_\sigma y^\rho_\nu)
\tag{5.17}
$$
(where $F_{\mu\nu} = A_{\mu,\nu} - A_{\nu,\mu}$, 
$({\frak g}_{\sigma\nu})$ denotes the Lorentz metric,
${\frak g}_{\sigma\nu} = 0$ for $\sigma \ne \nu$, 
$-{\frak g}_{11} = {\frak g}_{22} = {\frak g}_{33} = {\frak g}_{44}
= 1$,  
and $y^\sigma = {\frak g}^{\sigma\nu}A_\nu$), the standard regularity condition
(1.1) gives that $L$ is degenerate. For example, in the $n = 2$ case,
the De Donder momenta and Hamiltonian read
$$
\widetilde p^1_1 = \widetilde p^2_2 = 0, \quad
\widetilde p^1_2 = \widetilde p^2_1 = y^1_2 + y^2_1,
\quad \widetilde H = -\frac{1}{2} \left ( y^1_2 + y^2_1 \right )^2
  + \widetilde p^1_2 y^2_1 + \widetilde p^2_1 y^1_2.
\tag {5.18}
$$
Hence, momenta are not independent, and the corresponding Hamilton equations
must be treated as {\it constrained}. 
However, as we proved in Sec. 3, Lagrangian (5.17) is regularizable 
(and admits many global regularizations).
Let us choose one of them and compute the corresponding Hamiltonian and momenta.  
Put
$$
g^{\alpha \beta}_{\sigma \nu} = {{\pr^2 L} \over {\pr y^\sigma_\alpha \pr y^\nu_\beta}}
- {{\pr^2 L} \over {\pr y^\sigma_\beta \pr y^\nu_\alpha}}
\tag{5.19}
$$
(apparently, these $g$'s do not depend upon the $y^\sigma_i$'s, and satisfy 
(3.2), as desired).
The Lepagean equivalent (3.1) now reads,
$$
\rho = 
 L\omega_0 + {\pr L\over \pr y^\sigma_\alpha}\omega^\sigma \wedge \omega_\alpha +
\left ({{\pr^2 L} \over {\pr y^\sigma_\alpha \pr y^\nu_\beta}}
- {{\pr^2 L} \over {\pr y^\sigma_\beta \pr y^\nu_\alpha}} \right )
\omega^\sigma \wedge \omega^\nu \wedge \omega_{\alpha \beta},
\tag {5.20}
$$
and the regularity condition (3.3) leads to checking regularity of the following
matrix
$$
\left (4 \ {{\pr^2 L} \over {\pr y^\sigma_\beta \pr y^\nu_\alpha}}
- 3 \ {{\pr^2 L} \over {\pr y^\sigma_\alpha \pr y^\nu_\beta}} \right ).
\tag {5.21}
$$

\smallskip
(i) Let $X = R^2$.
Then we have $Y = R^2 \times R^2$, i.e. $m=2$, and the Lagrangian (5.17)
reads
$$
L = {1 \over 2}\left (y^1_2 + y^2_1 \right)^2.
$$
For the matrix (5.21) we obtain
$$
\pmatrix 
0&0&0&4 \\
0&1&-3&0\\
0&-3&1&0\\
4&0&0&0\\
\endpmatrix,
$$
i.e., it is {\it regular}. Consequently, the related Hamilton $p_2$-equations are 
{\it equivalent} with the Maxwell equations.
For the momenta we easily obtain 
$$
p^1_1 =  4 y^2_2, \quad 
p^1_2 =  - 3 y^1_2 + y^2_1 ,\quad  
p^2_2 =  4 y^1_1, \quad 
p^2_1 =\  y^1_2 -  3 y^2_1.
\tag{5.22}
$$
The inverse transformation to the Legendre transformation exists and takes the form
$$
y^1_1 =  {1 \over 4} p^2_2, \quad 
y^1_2 =  - {3 \over 8} p^1_2 - {1 \over 8} p^2_1 ,\quad 
y^2_2 =  {1 \over 4} p^1_1, \quad 
y^2_1 =\  - {1 \over 8} p^1_2 -  {3 \over 8} p^2_1.
$$
The Hamiltonian in the Legendre coordinates reads
$$
H = {1 \over 4}p^1_1 p^2_2 - {3 \over 8}p^1_2 p^2_1 -
  {1 \over 16}\left ( p^1_2 \right )^2 - {1 \over 16}\left (p^2_1 \right)^2.
\tag{5.23}
$$

\smallskip
(ii) Let $X = R^4$.
We have $m=4$, and the Lagrangian
(5.17) takes the form
$$
\aligned
L\, =&\, {1 \over 2}\left ( y^1_2\, +\, y^2_1 \right )^2\, +\,
{1 \over 2}\left ( y^1_3\, +\, y^3_1 \right )^2 \, +\,
{1 \over 2}\left ( y^1_4\, +\, y^4_1 \right )^2\\  \, &-\,
{1 \over 2}\left ( y^2_3\, -\, y^3_2 \right )^2 \, -\,
{1 \over 2}\left ( y^2_4\, -\, y^4_2 \right )^2\, - \,
{1 \over 2}\left ( y^3_4\, -\, y^4_3 \right )^2.
\endaligned
\tag{5.24}
$$
The matrix (5.21) becomes
$$
\pmatrix 
0&0&0&0&0&4&0&0&0&0&4&0&0&0&0&4& \\
0&1&0&0&-3&0&0&0&0&0&0&0&0&0&0&0&\\
0&0&1&0&0&0&0&0&-3&0&0&0&0&0&0&0&\\
0&0&0&1&0&0&0&0&0&0&0&0&-3&0&0&0&\\
0&-3&0&0&1&0&0&0&0&0&0&0&0&0&0&0&\\
4&0&0&0&0&0&0&0&0&0&4&0&0&0&0&4&\\
0&0&0&0&0&0&-1&0&0&-3&0&0&0&0&0&0&\\
0&0&0&0&0&0&0&-1&0&0&0&0&0&-3&0&0&\\
0&0&-3&0&0&0&0&0&1&0&0&0&0&0&0&0&\\
0&0&0&0&0&0&-3&0&0&-1&0&0&0&0&0&0&\\
4&0&0&0&0&4&0&0&0&0&0&0&0&0&0&4&\\
0&0&0&0&0&0&0&0&0&0&0&-1&0&0&-3&0&\\
0&0&0&-3&0&0&0&0&0&0&0&0&1&0&0&0&\\
0&0&0&0&0&0&0&-3&0&0&0&0&0&-1&0&0&\\
0&0&0&0&0&0&0&0&0&0&0&-3&0&0&-1&0&\\
4&0&0&0&0&4&0&0&0&0&4&0&0&0&0&0&\\
\endpmatrix,
$$
and one can easily check that it is regular.
For the momenta we get
$$
\aligned
p^1_1 &=  4 y^2_2 + 4 y^3_3 + 4 y^4_4, \quad
p^2_1 = y^1_2 - 3 y^2_1 ,\quad \ \  
p^3_1 = y^1_3 - 3 y^3_1,\, \quad \ \ 
p^4_1 = y^1_4 - 3 y^4_1,\, \\
p^2_2 &=  4 y^1_1 + 4 y^3_3 + 4 y^4_4, \quad 
p^1_2 =  y^2_1 - 3 y^1_2 ,\quad
p^3_2 = -y^2_3 - 3 y^3_2, \quad
p^4_2 = - y^2_4 - 3 y^4_2, \\
p^3_3 &=  4 y^1_1 + 4 y^2_2 + 4 y^4_4, \quad 
p^1_3 =  y^3_1 - 3 y^1_3 ,\quad
p^2_3 =  - y^3_2 - 3 y^2_3 , \quad
p^4_3 = - y^3_4 - 3 y^4_3, \\
p^4_4 &=  4 y^1_1 + 4 y^2_2 + 4 y^3_3, \quad 
p^1_4 =  y^4_1 - 3 y^1_4 ,\quad
p^2_4 =  - y^4_2 - 3 y^2_4 , \quad
p^3_4 =  - y^4_3 - 3 y^3_4, \\
\endaligned
\tag{5.25}
$$
and the Hamiltonian in the Legendre coordinates takes the form
$$
\aligned
H =& -\sfrac{1}{12}\Bigl (( p^1_1)^2 + (p^2_2)^2 + (p^3_3)^2 + (p^4_4)^2 
   - p^1_1p^2_2 - p^1_1p^3_3 - p^1_1p^4_4 - p^2_2p^3_3 - p^2_2p^4_4 - p^3_3p^4_4\Bigr )\\
&\, -\, \sfrac{1}{16}  \Bigl ( ( p^1_2)^2 + (p^2_1)^2 + (p^1_3)^2
 + (p^3_1)^2 + (p^1_4)^2 + (p^4_1)^2 \\
&\quad \quad  - ( p^2_3)^2 - (p^3_2)^2 - (p^2_4)^2
 - (p^4_2)^2 - (p^3_4)^2 - (p^4_3)^2 \Bigr )\\
&\, -\,
  \sfrac{3}{8}\Bigl ( p^1_2 p^2_1 + p^1_3 p^3_1 + p^1_4 p^4_1 + p^2_3 p^3_2
 + p^2_4 p^4_2 + p^3_4 p^4_3 \Bigr).
\endaligned
\tag{5.26}
$$

Let us compute the corresponding {\it satellite Lagrangian} for (5.17). Since
$$
g^{\alpha \beta}_{\sigma \nu} = \delta^\alpha_\nu\, \delta_\sigma^\beta -
\delta^\beta_\nu\, \delta_\sigma^\alpha,
\tag{5.27}
$$
we obtain
$$
l = 2( A^\mu_\nu A^\nu_\mu - A^\mu_\mu A^\nu_\nu) = 
2(\Tr ({A^\prime}^2) - (\Tr A^\prime)^2),
\tag{5.28}
$$
where $A^\prime$ denotes the matrix  $(\pr_\alpha A^\beta)$. 
Now, the dedonderization Lagrangian and the ``new" Hamiltonian for the 
electromagnetic field read
$$
\bar L = - \sfrac{1}{4} F_{\mu \nu}F^{\mu \nu}
+ 2( A^\mu_\mu A^\nu_\nu - A^\mu_\nu A^\nu_\mu), \quad
H = \widetilde H + 2( A^\mu_\mu A^\nu_\nu - A^\mu_\nu A^\nu_\mu).
\tag{5.29}
$$

\definition{Remark 2}
Note that the Lepagean equivalent $\rho_\lambda$ (2.7) is {\it not} an 
appropriate regularization of the electromagnetic field. Indeed,
in this case the matrix (3.3) takes the form
$$
\left (\ {{\pr^2 L} \over {\pr y^\sigma_\beta \pr y^\nu_\alpha}}
+ \ {{\pr^2 L} \over {\pr y^\sigma_\alpha \pr y^\nu_\beta}} \right ),
$$
and it is singular for the Lagrangian (5.17).
\enddefinition

\enddocument